\documentclass[preprint,fleqn]{revtex4}
\usepackage{bm}
\usepackage{amsmath}
\usepackage{float}
\usepackage{dcolumn}
\usepackage{amssymb}
\usepackage{graphics}
\usepackage{amsmath, amsfonts, amsthm, amssymb, graphicx,comment}
\usepackage{xcolor}

\def\be{\begin{equation}} \def\ee{\end{equation}}
\def\bea{\begin{eqnarray}} \def\eea{\end{eqnarray}}

\newcommand{\ud}{\mathrm{d}} \def \pd {\partial} 

\usepackage{color}

\begin{document}

\title{Dynamical analysis of the cosmology of mass-varying massive gravity }

\author{De-Jun Wu\footnote{Email: wudejun10@mails.ucas.ac.cn}}
\affiliation{School of Physics, University of Chinese Academy of
Sciences, Beijing 100049, China}

\author{Yifu Cai\footnote{Email: yifucai@physics.mcgill.ca}}
\affiliation{Department of Physics, McGill University, Montr\'eal, QC H3A 2T8, Canada}

\author{Yun-Song Piao\footnote{Email: yspiao@ucas.ac.cn}}
\affiliation{School of Physics, University of Chinese Academy of
Sciences, Beijing 100049, China}

\begin{abstract}
We study cosmological evolutions of the generalized model of
nonlinear massive gravity in which the graviton mass is given by a
rolling scalar field and is varying along time. By performing
dynamical analysis, we derive the critical points of this system
and study their stabilities. These critical points can be
classified into two categories depending on whether they are
identical with the traditional ones obtained in General
Relativity. We discuss the cosmological implication of relevant
critical points.

\end{abstract}

\maketitle

\section{Introduction}

Initiated by Fierz and Pauli (FP) \cite{Fierz:1939ix}, it has been
questioned for long time whether the graviton is allowed to
acquire a mass and leads to a consistent covariant modification of
General Relativity. At quadratic order the FP mass term is the
only ghost-free term describing a gravitational theory containing
five degrees of freedom \cite{VanNieuwenhuizen:1973fi}, but this
theory can not recover linearized Einstein gravity in the limit of
vanishing graviton mass, due to the existence of the van
Dam-Veltman-Zakharov (vDVZ) discontinuity arising from the
coupling between the longitude mode of the graviton and the trace
of the energy momentum tensor \cite{vanDam:1970vg,
Zakharov:1970cc}. It was later noticed that this troublesome mode
could be suppressed at macroscopic length scales due to the
so-called Vainshtein mechanism \cite{Vainshtein:1972sx}. However,
these nonlinear terms, which are responsible for the suppression
of vDVZ discontinuity, lead inevitably to the existence of the
Boulware-Deser (BD) ghost \cite{Boulware:1973my} and therefore
make the theory unstable \cite{ArkaniHamed:2002sp,
Creminelli:2005qk, Deffayet:2005ys, Gabadadze:2003jq}.

Recently, a family of nonlinear extension on the massive gravity theory was constructed by de Rham, Gabadadze and Tolley (dRGT) \cite{deRham:2010ik, deRham:2010kj}. In this model the BD ghosts can be removed in the decoupling limit to all orders in perturbation theory through a systematic construction of a covariant nonlinear action \cite{Hassan:2011vm, Hassan:2011hr, deRham:2011rn, deRham:2011qq} (see \cite{Hinterbichler:2011tt} for a review). As a consequence, the theoretical and phenomenological advantages of the dRGT model led to a wide investigation in the literature. For example, cosmological implications of the dRGT model are discussed in \cite{D'Amico:2011jj, Gumrukcuoglu:2011ew, Gumrukcuoglu:2011zh, DeFelice:2012mx, Gumrukcuoglu:2012aa, Koyama:2011wx, Comelli:2011zm, Crisostomi:2012db, Cardone:2012qq, Gratia:2012wt, Kobayashi:2012fz, D'Amico:2012pi, Fasiello:2012rw, D'Amico:2012zv, Langlois:2012hk, Gong:2012yv, Huang:2012pe, Saridakis:2012jy, Cai:2012ag, Rham:2010ri, Akrami:2012ri,YlZhang:2012}; black holes and spherically symmetric solutions were analyzed in \cite{Koyama:2011xz, Koyama:2011yg, Sbisa:2012zk, Nieuwenhuizen:2011sq, Gruzinov:2011mm, Comelli:2011wq, Berezhiani:2011mt, Sjors:2011iv, Chkareuli:2011te, Brihaye:2011aa, Cai:2012db}; and connections to bi-metric gravity models were studied in \cite{Damour:2002wu, Hassan:2011tf, Hassan:2011zd, vonStrauss:2011mq, Hassan:2011ea, Hassan:2012wr, Volkov:2011an, Volkov:2012wp, Paulos:2012xe, Hinterbichler:2012cn, Baccetti:2012bk, Baccetti:2012re, Baccetti:2012ge, Berg:2012kn}.

Among these phenomenological studies, a generalized version of the
dRGT model was constructed in Ref. \cite{Huang:2012pe} that the
graviton mass can be determined by a rolling scalar field. This
mass-varying massive gravity (MVMG) model was argued to be free of
the BD ghosts as well through examining the constraint system in
the Hamiltonian formulation. Thus, it is interesting to
investigate the cosmological implications of this model,
especially its late time evolution. In the present work we perform
a phase-space and stability analysis of this model, and
investigate the possible cosmological dynamics in a systematical
way. This method was widely applied in the literature and was
proven to be very powerful particularly in the study of dark
energy physics \cite{Halliwell:1986ja, Ferreira:1997au,
Copeland:1997et, Guo:200304, Guo:200306,Guo:2004fq, Gong:2006sp,
Cai:2006dm, Zhou:2007, Chen:2008ft, Zhou:2009, Leon:2012mt} (see
also Refs. \cite{Copeland:2006wr, Cai:2009zp} for relevant
reviews). In the present model the cosmological system shows a
couple of critical points at late times, and they can generally be
classified into two categories depending on whether they are
identical with the traditional ones obtained in General Relativity
(GR). Although some of the critical points are able to be
recovered in GR in the limit of vanishing graviton mass, their
background dynamics are different since of the graviton potential.
By performing stability analysis we find that the parameter space
of the model is tightly constrained and the stable cosmological
evolutions are quite trivial. Moreover, there exist some new
critical points in the MVMG model which might be of theoretical
interests, but the corresponding cosmological evolutions are
basically ruled out by observations.

The present paper is organized as follows. In Section II, we briefly review the MVMG model and its cosmological equations of motion. Then we perform a detailed phase-space and stability analysis of this cosmological system and summarize the results in Section III. Finally, we conclude with a discussion in Section IV.

\section{The MVMG Cosmology}

To begin with, we briefly review the MVMG model constructed in \cite{Huang:2012pe}. This model requires the graviton mass to a function of another scalar field by introducing a nontrivial potential term $V(\psi)$ and thus it is able to vary along cosmic evolution. General Relativity can be automatically recovered when this scalar field $\psi$ sits at $V(\psi)=0$. In order to better control dynamics of the scalar field, the model also includes an additional potential $W(\psi)$. Therefore, the complete action can be expressed as
\begin{equation}
 S = \int {{d^4}x} \sqrt { - g} \left[ {\frac{{M_P^2}}{2}R + V\left( \psi  \right)\left( {{U_2} + {\alpha _3}{U_3} + {\alpha _4}{U_4}} \right) - \frac{1}{2}{\partial _\mu }\psi {\partial ^\mu }\psi  - W\left( \psi  \right)} \right]~,
\end{equation}
The ${U_2}$, $U_3$ and $U_4$ terms are the graviton potentials following the dRGT model, which are given by,
\begin{align}
 U_2  =  \mathcal{K}^\mu_{[\mu}\mathcal{K}^\nu_{\nu]} ~, \quad
 U_3  = \mathcal{K}^\mu_{[\mu}\mathcal{K}^\nu_{\nu}\mathcal{K}^\rho_{\rho]} ~, \quad
 U_4  = \mathcal{K}^\mu_{[\mu}\mathcal{K}^\nu_{\nu}\mathcal{K}^\rho_{\rho}\mathcal{K}^\sigma_{\sigma]} ~,
\end{align}
with
\begin{equation}
 \mathcal{K}^\mu_\nu =
 \delta^\mu_\nu-\sqrt{g^{\mu\rho}f_{AB}\pd_\rho \phi^A
 \pd_\nu \phi^B } ~.
\end{equation}
Moreover, $f_{AB}$ is a fiducial metric, which is often chosen as Minkowski $f_{AB}=\eta_{AB}$. The four $\phi^A(x)$ are St\"{u}ckelberg scalars introduced to restore general covariance under
\be
 g_{\mu\nu}(x) \to \frac{\pd x^\rho}{\pd x'^\mu} \frac{\pd x^\sigma}{\pd x'^\nu} g_{\rho\sigma}(x) ~, \quad
 \phi^A(x) \to \phi^A(x)~; \qquad
  x^\mu \to x'^{\mu}~.
\ee
For the case $f_{AB}=\eta_{AB}$, the St\"{u}ckelberg scalars form Lorentz 4-vectors in the internal space. By performing the Hamiltonian constraint of the system it is argued that this model is still free of BD ghosts \cite{Huang:2012pe}.

In the following we take into account the regular matter component in the total action which minimally coupled to the gravitational system. Further, we consider the fiducial metric to be Minkowski
\be
 f_{AB}=\eta_{AB},
\ee
and assume that the dynamical and fiducial metrics are of diagonal forms simultaneously for simplicity. Specifically, we consider a flat Friedmann-Robertson-Walker (FRW) metric
\begin{align}
 \ud^2 s &= -N(\tau)^2 \ud \tau^2 +a(\tau)^2 \delta_{ij} \ud x^i \ud x^j,
\end{align}
and for the St\"{u}ckelberg fields we choose the ansatz
\begin{eqnarray}
 \phi^0 = b(\tau)  ,  ~~~~\phi^i = a_0 x^i ,
\end{eqnarray}
where $a_0$ is constant.

Finally, the total action for the cosmological background reduces to
\be
 S_T = \int \ud^4 x \left[
-3M_P^2  \frac{a'^2a}{N} + V(\psi) (u_2^F +\alpha_3 u_3^F +
\alpha_4 u_4^F)  +\frac{a^3}{2N} \psi'^2 -Na^3W(\psi)\right]
+ S_m   ,
\ee
where
\begin{align}
 u_2^F & = 3a(a-a_0)(2Na-a_0 N-ab')  ,  \\
 u_3^F & = (a-a_0)^2 (4Na-a_0 N-3ab')  ,  \\
 u_4^F & = (a-a_0)^3(N-b')  ,
\end{align}
and we have define ${~~}'=\ud /\ud \tau$. Variations of the total action $S_T$ with respect to $N$ and $a$ lead to the two Friedmann equations
\begin{eqnarray}
 3M_P^2 H^2& =& \rho_{MG}+\rho_m     ,\\
 -2 M_P^2 \dot{H}& =&\rho_{MG}+p_{MG}+\rho_m +p_m  ~,
\end{eqnarray}
respectively. In the above equations, $\rho_m$ and $p_m$ are the density and pressure for the matter component, respectively. The effective density and pressure are given by
\begin{align}
 \rho_{MG} &=\frac12 \dot{\psi}^2+W(\psi)+V(\psi)(X-1)f_3(\alpha_i,X)+V(\psi) (X-1)f_1(\alpha_i,X) , \\
 p_{MG} &=\frac12 \dot{\psi}^2-W(\psi)- V(\psi)(X-1)f_3(\alpha_i,X) -V(\psi)(\dot{b}-1) f_1(\alpha_i,X),
\end{align}
with
\begin{align}
 f_1(\alpha_i,X) &= (3-2X)+\alpha_3(3-X)(1-X)+\alpha_4(1-X)^2 , \\
 f_2(\alpha_i,X) &= (1-X)+\alpha_3(1-X)^2+\frac{\alpha_4}{3}(1-X)^3 , \\
 f_3(\alpha_i,X) &=(3-X)+\alpha_3(1-X),
\end{align}
and $X \equiv {a_0}/{a}$.

Moreover, variations of the total action with respect to $b$ and $\psi$ give rise to the following two equations of motion,
\begin{eqnarray}
\label{EoM_b}
&& V(\psi)Hf_1(\alpha_i,X)+\dot{V}(\psi)f_2(\alpha_i,X)  = 0 ~, \\
\label{EoM_psi}
&& \ddot{\psi}+3H\dot{\psi}+\frac{\ud W}{\ud \psi}+\frac{\ud V}{\ud \psi}[(X-1) (f_3(\alpha_i,X)+f_1(\alpha_i,X)) +3\dot{b}f_2(\alpha_i,X)]  =0~.
\end{eqnarray}
which will be frequently used in detailed calculations later. In addition, it is convenient to define an effective equation of state parameter for modified gravity terms as follows,
\be
 {w_{MG}} = \frac{{{\rho _{MG}}}}{{{p_{MG}}}} = \frac{{\frac{1}{2}{{\dot \psi }^2} + W(\psi ) + V(\psi )(X - 1){f_3}({\alpha _i},X) + V(\psi )(X - 1){f_1}({\alpha _i},X)}}{{\frac{1}{2}{{\dot \psi }^2} - W(\psi ) - V(\psi )(X - 1){f_3}({\alpha _i},X) - V(\psi )(\dot b - 1){f_1}({\alpha _i},X)}}~,
\ee
and the total equation of state parameter is defined as:
\be
 {w_{tot}} = \frac{{{\rho _{MG}} + {\rho _m}}}{{{p_{MG}} + {p_m}}} =  - 1 - \frac{{2\dot H}}{{3{H^2}}} ~.
\ee

\section{Dynamical Framework of MVMG cosmology}

In this section we perform a detailed phase-space analysis of cosmic evolutions described by the MVMG model. Following the method extensively developed in \cite{Halliwell:1986ja, Ferreira:1997au, Copeland:1997et, Guo:200304, Guo:200306, Guo:2004fq, Gong:2006sp, Cai:2006dm, Zhou:2007, Chen:2008ft, Zhou:2009} (see also \cite{Leon:2012mt} for a recent analysis in the frame of generalized Galileon cosmology), we first transform the dynamical system into the autonomous form.

\subsection{Dynamics of the autonomous system}

In general, for a dynamical system one can suitably choose a group of auxiliary variables, and the corresponding equation of motion can be expressed as a group of first-order differential equations, respectively. Namely, to illustrate the method of phase-space analysis, we consider the following two-variable dynamical system:
\be\label{example}
 \dot x = f(x,y)~, ~~\dot y = g(x,y) ~.
\ee
The system is said to be autonomous if $f$ and $g$ do not contain explicit time-dependent terms. A point $({x_c},{y_c})$ is said to be a critical point of the autonomous system if $(f,g){|_{({x_c},{y_c})}} = 0$. One can check whether the system approaches one of the critical points or not by performing the stability analysis around the fixed points. Specifically, one can introduce $\delta x$ and $\delta y$ as small perturbations and expand the differential equations (\ref{example}) to the first order of $\delta x$ and $\delta y$ around the critical point, and then can derive out the following equations of motion,
\be\label{EoM_autonomous}
 \frac{d}{{dN}}\left( {\begin{array}{*{20}{c}}
 {\delta x}\\
 {\delta y}
 \end{array}} \right) = {\left( {\begin{array}{*{20}{c}}
 {\frac{{\partial f}}{{\partial x}}}&{\frac{{\partial f}}{{\partial y}}}\\
 {\frac{{\partial g}}{{\partial x}}}&{\frac{{\partial g}}{{\partial y}}}
 \end{array}} \right)_{\left( {x = {x_c},y = {y_c}} \right)}}\left( {\begin{array}{*{20}{c}}
 {\delta x}\\
 {\delta y}
 \end{array}} \right)~.
\ee
As a consequence, the general solution for the evolution of linear perturbations can be written as
\be
\delta x = {c_1}{e^{{\mu _1}N}} + {c_2}{e^{{\mu _2}N}},
\ee
\be
\delta y = {c_3}{e^{{\mu _1}N}} + {c_4}{e^{{\mu _2}N}},
\ee
where $\mu _1 $ and $\mu _2 $ are the two eigenvalues of matrix in the left hand side of Eq. \eqref{EoM_autonomous},
and $c_1$, $c_2$, $c_3$ and $c_4$ are constant coefficients. If $\mu _1<0 $ and $\mu _2<0 $, then the point is stable, which means the system could evolve to this fixed point eventually. The method can be extended to a system with many variables, a critical point is stable if  the real parts of all the corresponding  eigenvalues are negative.

In the model we consider, there are five dimensionless variables
\bea
{x_\rho } = \frac{{\sqrt {{\rho _m}} }}{{\sqrt 3 {M_p}H}}
,\quad
{x_\psi } = \frac{{\dot \psi }}{{\sqrt 6 {M_p}H}}
,\quad
{x_W} = \frac{{\sqrt {W(\psi) }}}{{\sqrt 3 {M_p}H}}
,\quad
{x_V} = \frac{{\sqrt{ V(\psi)}}}{{\sqrt 3 {M_p}H}}
,\quad
{x_a} = \frac{a_0}{a}~.
\eea
Among them, $x_a$ and $x_\rho$ can be determined by background equations of motion as will be analyzed in this subsection. Thus the system only involves three independent variables.

Making use of these variables, one can reexpress the Friedmann equation as follows,
\begin{equation}
 1 = {x_\rho }^2 + {x_\psi }^2 + x_W^2 + {x_V}^2({x_a} - 1)({f_1}({x_a}) + {f_3}({x_a}))~.
\end{equation}
which now is a constraint equation.

Then, we particularly choose an exponential potential
\be
 W\left( \psi  \right) = \exp \left[ { - \frac{\lambda }{{{M_p}}}\psi } \right]~,
\ee
with $\lambda>0 $. Moreover, we parameterize the form of $b$ as a linear function of cosmic time, which is given by
\be
 b=Bt ~~{\rm with}~~ B>0 ~,
\ee
so that this autonomous system is analytically solvable. In addition, we also assume the matter fluid satisfies a barotropic equation of state ${p_m} = \left( {\gamma  - 1} \right){\rho _m}$, with $\gamma$ being a constant and $0 <\gamma\leq2$. From equation \eqref{EoM_b}, one can get
\be
 \dot V\left( \psi  \right) = \frac{{dV}}{{d\psi }}\dot \psi  =  - \frac{{V\left( \psi  \right)H{f_1}\left( {{x_a}} \right)}}{{{f_2}\left( {{x_a}} \right)}},
\ee
and thus
\be
 \frac{{dV}}{{d\psi }} =  - \frac{{VH{f_1}\left( {{x_a}} \right)}}{{\dot \psi {f_2}\left( {{x_a}} \right)}}~.
\ee
if $\dot\psi\ne0$. Note that, if $\dot\psi=0$ the mass term $V(\psi)$ is fixed and then the model would reduce to the dRGT version which has been shown in \cite{D'Amico:2011jj} that a flat FRW background is not allowed. Therefore, we will not consider this case in the present work. Moreover, using the auxiliary variables, one can transform the equations of motion to the following autonomous forms,
\begin{eqnarray}
 \frac{1}{H}\frac{d}{{dt}}{x_\rho } &=& \frac{3}{2}{x_\rho }\left( {\gamma x_\rho ^2 + 2x_{^\psi }^2 + {f_1}\left( {{x_a}} \right)\left( {{x_a} - B} \right)x_V^2} \right) - \frac{3}{2}\gamma {x_\rho } ~,\\
 \frac{1}{H}\frac{d}{{dt}}{x_\psi } &=& \frac{{3{x_\psi}}}{2}(\gamma x_\rho ^2 + 2x_\psi ^2 + {f_1}({x_a})x_V^2({x_a}- B)) - 3{x_\psi } + \frac{{\lambda\sqrt 6 }}{2}x_W^2 \nonumber\\
   && + \frac{{x_V^2{f_1}({x_a})}}{{2{x_\psi}{f_2}({x_a})}}(({x_a} - 1)({f_3}({x_a}) + {f_1}({x_a})) + 3B{f_2}({x_a})) ~,\\
 \frac{1}{H}\frac{d}{{dt}}{x_W} &=& \frac{{3{x_W}}}{2}(\gamma x_\rho ^2 + 2x_\psi ^2 + {f_1}({x_a})x_V^2({x_a} - B)) - \frac{{\lambda\sqrt 6 }}{2}{x_\psi }{x_W} ~,\\
 \frac{1}{H}\frac{d}{{dt}}{x_V} &=& \frac{{3{x_V}}}{2}(\gamma x_\rho ^2 + 2x_\psi ^2 + {f_1}({x_a})x_V^2({x_a} -  B)) - \frac{{{f_1}({x_a}){x_V}}}{{2{f_2}({x_a})}} ~,\\
\label{eq_x_a}
 \frac{1}{H}\frac{d}{{dt}}{x_a} &=&  - {x_a} ~.
\end{eqnarray}

We restrict our discussion of the existence and stability of critical points to the expanding universes with $H > 0$. The critical points correspond to those fixed points where $\frac{d}{{Hdt}}{x_\rho } = 0, \frac{d}{{Hdt}}{x_\psi } = 0, \frac{d}{{Hdt}}{x_W} = 0, \frac{d}{{Hdt}}{x_V} = 0$ and $\frac{d}{{Hdt}}{x_a} = 0$, and there are self-similar solutions satisfying
\be
 \frac{{\dot H}}{{{H^2}}} = -\frac{3}{2}(\gamma x_\rho ^2 + 2x_\psi ^2 + {f_1}({x_a})x_V^2({x_a} - B)).
\ee

The equation (\ref{eq_x_a}) simply suggests $x_a=0$. It can be
shown  that having the variable $x_a$ in the system always bring a
eigenvalue of $-1$ and leave the other eigenvalues unchanged. Thus
in the following, for simplicity we will take $x_a=0$, which dose
not affect the final results.

By defining \bea{f_1} = 3 + 3{\alpha _3} + {a_4},{f_2} =
\frac{{{f_1}}}{3},{f_3} = 3 + {\alpha _3},\eea and using $x_a=0$,
the constrain equation reduces to
\begin{equation}
1 = {x_\rho }^2 + {x_\psi }^2 + x_W^2 - {x_V}^2({f_1} + {f_3}).
\end{equation}
Further, we use the constrain equation to eliminate the variable ${x_\rho}$ and get
\bea
{{x_\psi }}(\gamma (1 - x_W^2) + (2 - \gamma )x_\psi ^2 + c{x_V}^2) - 2{x_\psi } + \frac{{\lambda \sqrt 6 }}{3}x_W^2 - d\frac{x_V^2}{{{x_\psi }}}= 0~,
\eea
\begin{equation}
{3{x_W}}(\gamma (1 - x_W^2) + (2 - \gamma )x_\psi ^2 + {c}x_V^2) - {\lambda\sqrt 6 }{x_\psi }{x_W} = 0~,
\end{equation}
\begin{equation}
{x_V}(\gamma (1 - x_W^2) + (2 - \gamma )x_\psi ^2 + {c}x_V^2) - {x_V} = 0~,
\end{equation}
and the self-similar solutions reduce to
\be\label{selfsimilar}
\frac{{\dot H}}{{{H^2}}} = -\frac{3}{2}(\gamma (1 - x_W^2) + (2 - \gamma )x_\psi ^2 + c{x_V^2})~,
\ee
where in order for convenience we have defined two dimensionless parameters
$c=({f_1} + {f_3})\gamma  - B{f_1}$
and
$d= {f_1}(1-B) + {f_3}$.

\subsection{Phase-space analysis and results}

We summarize the fixed points of this autonomous system and their stability analysis in Table I and Table II, respectively. In the following we discuss these solutions case by case.

\begin{table}
\begin{tabular}{cccc}
\hline&Points that can be recovered in GR&\\
\hline
\hline
Lable &${x_\psi }$ & ${x_W}$& ~~~~~~~~${x_V}$\\ \hline
(a1)&1 &0 &~~~~~~~~0 \\
(a2)&-1 &0 &~~~~~~~~0 \\
(b)&$\frac{\lambda }{{\sqrt 6 }}$ &$\sqrt {1 - \frac{{{\lambda ^2}}}{6}}$ &~~~~~~~~$0$\\
(c)&$ \sqrt {\frac{3}{2}}\frac{{ \gamma }}{\lambda }$ &$\sqrt {\frac{3}{2}}\frac{{ \sqrt {  (  2 - \gamma )\gamma } }}{\lambda }$ &~~~~~~~~$0$ \\
\hline
&Points that can't be recovered in GR&\\
\hline\hline
Lable &${x_\psi }$ & ${x_W}$& ~~~~~~~${x_V}$\\ \hline
(d)&$\sqrt {\frac{3}{2}}\frac{ 1 }{\lambda }$ &$\sqrt {x_\psi ^2 + dx_V^2} $ &~~~~~~~~$\sqrt {\frac{{\left( {\gamma  - 1} \right)}}{{c - d\gamma }}\left( {\frac{3}{{{\lambda ^2}}} - 1} \right)} $\\

(e1)&$\sqrt\frac{{  { d(\gamma- 1  ) } }}{{ {c + d( \gamma- 2  )} }}$ &$0$ &~~~~~~~~$\sqrt\frac{{{1 - \gamma } }}{{ {c + d( \gamma- 2  )} }}$\\
(e2)&$-\sqrt\frac{{  { d(\gamma- 1 ) } }}{{ {c + d( \gamma- 2  )} }}$ &$0$ &~~~~~~~~$\sqrt\frac{{ {1 - \gamma } }}{{ {c + d(\gamma - 2  )} }}$
\\ \hline

\end{tabular}
\caption{The critical points in a spatially flat FRW universe with exponential potentials in the MVMG cosmology.}
\end{table}

\begin{table}
\begin{tabular}{cccc}
\hline&Points that can be recovered in GR&\\
\hline\hline
Lable & Existence& Stability& Equation of state\\  \hline

(a1) &all $\gamma$ and $\lambda $& unstable&1\\
(a2) &all $\gamma$ and $\lambda $& unstable&1\\
(b)&${\lambda ^2} < 6$ & stable if ${\lambda ^2} < \min(3\gamma,3)$&$-1+\frac{\lambda^2}{3}$\\
(c) &${\lambda ^2} > 3\gamma $ &stable if $\gamma < 1 $&$-1+\gamma$\\
\hline
&Points that can't be recovered in GR&\\
\hline\hline
Lable & Existence& Stability& Equation of state\\  \hline
(d)&$\frac{{\left( {\gamma  - 1} \right)}}{{c - d\gamma }}\left( {\frac{3}{{{\lambda ^2}}} - 1} \right) > 0,d >  - \frac{{x_\psi ^2}}{{x_V^2}}$  & ~~~~~stable if $\gamma  > 1,{\lambda ^2} >3 ,E>0$&0\\
(e1)& $d<0,\frac{{1 - \gamma }}{{c + d(  \gamma - 2 )}} > 0$  &~~~~~stable if $\gamma  > 1, {\lambda ^2} > \frac{3}{2}\left( {1 + \frac{{c - d}}{{d\left( {\gamma  - 1} \right)}}} \right)$&0\\
(e2)&  $d<0,\frac{{1 - \gamma }}{{c + d(  \gamma- 2  )}} > 0$  &unstable&0\\
 \hline
\end{tabular}
\caption{The properties of the critical points. The definitions of $E$ is given in the text below. }
\end{table}

From Table I, one may notice that the points  are either of $x_{V}=0$ or not. For those with $x_V=0$ (which are the points (a1), (a2), (b), and (c), respectively), the final state of the universe corresponds to that the background evolution is mainly determined by the exponential potential instead of the graviton mass. These solutions are able to be recovered in the limit of vanishing graviton mass and thus are consistent with the standard results obtained in GR. A further stability analysis suggests the points (a1) and (a2) are unstable. For the point (b), the solution is allowed if $\lambda^2<6$ and stable against perturbations when ${\lambda ^2} < \min(3\gamma,3)$. For the point (c), the solution exists when $\lambda^2>3\gamma$ but is stable only if $\gamma<1$. We list the eigenvalues of these solutions in detail as follows.

Point (a1) is a kinetic-dominated solution. The linearization of the system yields three eigenvalues
\be
 m_1=\frac{3}{2}, m_2=6 - 3\gamma ,m_3=3-\sqrt {\frac{3}{2}} \lambda ~.
\ee
The corresponding effective equation of state for the whole system approaches to $w=1$ which implies a stiff fluid dominant phase at late times of the universe.

Point (a2) is a kinetic-dominated solution as well. The linearization of the system yields three eigenvalues
\be
 m_1=\frac{3}{2},   m_2=6 - 3\gamma ,m_3=3+\sqrt {\frac{3}{2}} \lambda ~,
\ee
and correspondingly, the destiny of the universe is the same as Point (a1).

Point (b) is a scalar-dominated solution since ${x_\psi}^2+{x_W}^2 = 1$, and only exists when ${\lambda ^2} < 6$. The linearization of the system yields three eigenvalues
\be
 m_1=\frac{1}{2}( {\lambda ^2}- 3  )~, ~~m_2=\frac{1}{2}( {\lambda ^2}- 6  )~, ~~m_3={\lambda ^2}- 3\gamma~.
\ee
The corresponding total effective equation of state is $w_{tot}=-1+\frac{\lambda^2}{3}$ which is always less than unity.

Point (c) is a solution depending on both the scalar field and the matter fluid, since ${x_\psi}^2+{x_W}^2 = \frac{{3\gamma }}{{{\lambda ^2}}}$. The linearization of the system yields three eigenvalues
\begin{eqnarray}
 m_1&=&\frac{3}{2}(  \gamma- 1  ) ~,\nonumber\\
 m_2&=& - \frac{{3\left( {2 - \gamma } \right)}}{4}\left( {1 + \sqrt {1 - \frac{{8\gamma \left( {{\lambda ^2} - 3\gamma } \right)}}{{{\lambda ^2}\left( {2 - \gamma } \right)}}} } \right)~,\nonumber\\
 m_3&=& - \frac{{3\left( {2 - \gamma } \right)}}{4}\left( {1 - \sqrt {1 - \frac{{8\gamma \left( {{\lambda ^2} - 3\gamma } \right)}}{{{\lambda ^2}\left( {2 - \gamma } \right)}}} } \right) ~.
\end{eqnarray}
The total effective equation of state is given by $w_{tot}=-1+\gamma$ in this solution.

Point (d) is a solution depending on both the scalar field and the graviton mass since $x_\psi ^2 + x_W^2 - x_V^2({f_1} + {f_3}) = 1$, the linearization of the system yields three eigenvalues
\begin{eqnarray}
 m_1&=&3 (1- \gamma)~,\nonumber\\
 m_2&=&- E + \sqrt {3\left( {3 - {\lambda ^2}} \right)x_W^2 + {E^2}}
~,\nonumber\\
 m_3&=& - E - \sqrt {3\left( {3 - {\lambda ^2}} \right)x_W^2 + {E^2}} ~,
\end{eqnarray}
where $E$ is a dimensionless parameter with $E= \frac{3}{4} + \frac{{d\left( { 3-{\lambda ^2} } \right)\left( {1 - \gamma } \right)}}{{2\left( {c - d\gamma } \right)}}.$

Point (e1) is a solution determined by the kinetic term of the scalar field and the graviton mass since $x_W= 0$ at late times. Its linearization yields three eigenvalues
\begin{eqnarray}
 m_1&=&-3~,\nonumber\\
 m_{2}&=&3 (1- \gamma) ~,\nonumber\\
 m_{3}&=&\frac{3}{2} - \lambda  \sqrt {\frac{3}{2}}\sqrt \frac{{ d (\gamma - 1) }}{{{c + d(  \gamma- 2 )} }}~.
\end{eqnarray}

Point (e2) is similar to Point (e1) with its eigenvalues given by
\begin{eqnarray}
 m_1&=&-3~,\nonumber\\
 m_{2}&=&3 (1- \gamma) ~,\nonumber\\
 m_{3}&=&\frac{3}{2} +\lambda  \sqrt {\frac{3}{2}}\sqrt \frac{{ d (\gamma - 1) }}{{{c + d(  \gamma- 2  )} }}~.
\end{eqnarray}
It is clear that $m_{3}$ is positive, so the solution is unstable.

It is worth noticing that the total effective equation of state of Points (a1), (a2), (b), and (c) are the same as the massless case. This result shows that the effect of graviton mass does not contribute manifestly in the solutions corresponding to GR. Moreover, if we substitute the critical points (d), (e1) and (e2) into the Eq. \eqref{selfsimilar}, then we get $\dot{H}/H^2 = -3/2$ which implies a matter domination at late times. This conclusion suggests that the appearance of a scalar field dependent graviton mass could strongly fix the background dynamics of the universe which manifestly conflicts with the observational fact of late time acceleration.

We would like to point out that the existence of points (d), (e1) and (e2) does not conflict with the fact that massive gravity theory does not allow flat FRW cosmologies, the fixed value of ${x_V}$ does not mean that the graviton mass is fixed. What actually happens is that the graviton mass and $H$ both gradually approach 0 as $t\rightarrow\infty $ with the magnitude about $1/{t^2}$ and $1/t$ respectively.

Finally, we can conclude that if a MVMG model is of cosmological interest, then its model parameters have to satisfy either ${\lambda ^2} < \min(3\gamma,3)$ (required by the stability of Point (b)) or $\gamma<1$ (required by the stability of Point (c)). Specifically, the solution of Point (b) corresponds to that the final evolution of the universe is determined by the scalar field and the effect of graviton mass is totally negligible, and thus this solution is quite trivial. Moreover, the solution of Point (c) corresponds to that the destiny of the universe is determined by the combined effects of the scalar field and the matter fluid. However, the stability of Point (c) requires the barotropic equation of state of matter fluid to be less than unity, which implies the corresponding matter fluid has to violate the strong energy condition and thus obviously conflict with cosmological observations.

\section{Conclusions}

In the present paper we have studied the dynamical behavior of the
MVMG cosmology. This model, due to the varying of the graviton
mass, possesses plentiful phenomenological properties and
consequently has attracted many attentions in the literature.
Namely, it was shown to be able to violate the null energy
condition effectively and thus could be applied to realize the
phantom divide crossing \cite{Saridakis:2012jy}; further, it was
also applied into early universe and a class of bouncing and
oscillating solutions were reconstructed \cite{Cai:2012ag}.
However, while phenomenological studies of this model is still
proceeding extensively, it is necessary to investigate the phase
space of the model and examine the stability of the critical
points existing in cosmological trajectories. Thus we performed a
detailed dynamical analysis of the MVMG model with a specifically
chosen potential for the cosmic scalar field, namely an
exponential potential. Our result reveals that there are mainly
two types of critical points in this model. One type of critical
points correspond to the case that they can recover the standard
results in GR if the graviton mass is chosen to be vanishing; the
other type of critical points then is discontinuous with GR in the
massless limit. We analyze these points respectively and find that
there are only two critical points which might be stable against
perturbations and both two belong to the first type. However, one
of these two solutions  is difficult to accommodate with
cosmological observations since its stability requires the matter
components in the universe to violate strong energy condition.
Eventually, there is only one viable solution in this model but
the final state of the universe is completely determined by the
cosmic scalar field and there is no effect of modified gravity. As
a consequence, the MVMG cosmology severely degenerate with the
standard cosmology based on GR.

We would like to point out that, although the graviton mass does
not give rise to observable effects on the background evolution at
late times, it may still leave signatures on cosmological
perturbations and thus affect the large scale structure formation.
Moreover, in our investigation, we focused on a particularly
chosen potential for the cosmic scalar field. It would be
interesting to generalize the case to a much more generic
potential and verify if the phase space of the viable solutions
could be enlarged.

{\bf{Note added}}: While this work was being finalized, we learned of a related work by G.~Leon, E.~N.~Saridakis and J.~Saavedra which will be appeared on arXiv \cite{Leon:2013}. Part of their content overlaps with ours, and their conclusions are similar as well while their focus is on a generalized structure of the phase space.

\section*{Acknowledgements}
We would like to thank E.~N.~Saridakis and S.~Y. Zhou for
discussion and comments, and also S. Y. Zhou for his initial
collaboration on this project. The work of DJW and YSP is
supported in part by NSFC under Grant No:11075205, 11222546, in
part by the Scientific Research Fund of GUCAS (NO:055101BM03), and
in part by National Basic Research Program of China under
No:2010CB832804. The work of YC is supported by an NSERC Discovery
Grant.

\end{document}